\documentclass[usenatbib]{jaa}
\usepackage{epsfig}
\usepackage{float}
\usepackage{amsmath}
\usepackage{rotating}
\usepackage{natbib}
\usepackage{deluxetable}
\usepackage[T1]{fontenc}
\usepackage[british]{babel}
\usepackage[varg]{txfonts}

\begin{document}
\title[Differential Light Curves of AGN]
{On the photometric error calibration for the `differential light curves' of point-like Active Galactic Nuclei}
\author[Goyal et al.]
       {Arti Goyal,$^{1,2}$$\thanks{E-mail: arti.aries@gmail.com}$
        Mukul Mhaskey,$^{3}$ Gopal-Krishna,$^{1,2}$ Paul J. Wiita,$^{4}$
\newauthor C. S. Stalin$^{5}$ and Ram Sagar,$^{6}$ \\
$^{1}$ National Centre for Radio Astrophysics (NCRA), TIFR, Pune 411 007, India. \\
$^{2}$ Inter University Centre for Astronomy and Astrophyics (IUCAA), Post Bag 4 Ganeshkhind, Pune University Campus, Pune 411 007, India \\ 
$^{3}$ Department of Physics, Pune University, Pune 411 007, India. \\
$^{4}$ Department of Physics, The College of New Jersey, PO Box 7718, Ewing, NJ 08628-0718, USA. \\
$^{5}$ Indian Institute of Astrophysics (IIA) Bangalore 560 034, India. \\
$^{6}$ Aryabhatta Research Institute of Observational Sciences (ARIES), Manora Peak, Naini Tal 263 129, India. \\
}
\date{Received \today; accepted \today}
\pubyear{xxxx}
\volume{xx}
\date{Received xxx; accepted xxx}
\maketitle
\label{firstpage}
\begin{abstract}
It is important to quantify the underestimation of rms photometric 
errors returned by the commonly used $\emph APPHOT$ algorithm 
in the $\emph IRAF$ software, in the context of differential 
photometry of point-like AGN, because of the crucial 
role it plays in evaluating their variability properties. 
Published values of the underestimation factor, $\eta$,
using several different telescopes, 
lie in the range 1.3 - 1.75. The present study aims to revisit 
this question by employing an exceptionally large data set of 
262 differential light curves (DLCs) derived from 262 pairs 
of non-varying stars monitored under our ARIES AGN monitoring 
program for characterizing the intra-night optical variability 
(INOV) of prominent AGN classes. 
The bulk of these data were taken with the 1-m Sampurnanad Telescope (ST).
We find $\eta$ = 1.54$\pm$0.05 
which is close to our recently reported value of $\eta$ = 1.5. 
Moreover, this consistency holds at least up to a brightness 
mismatch of 1.5 mag between the paired stars. From this 
we infer that a magnitude difference of at least up to 1.5 
mag between a point-like AGN and comparison star(s) monitored 
simultaneously is within the same CCD chip acceptable, 
as it should not lead to spurious claims of INOV. 
\end{abstract}
\section{Introduction}
\label{intro}
Observations of intensity variations at different wavelengths
constitute a highly effective probe of the physics of active
galactic nuclei (AGN). In the optical domain, numerous such
studies have been carried out, covering time scales down to hours 
and even minutes, sometimes coordinated with monitoring in other
wavebands (e.g., \citealt*{1989Natur.337..627Miller}; \citealt*{1995ARA&A..33..163Wagner}; 
\citealt{1995ApJ...452..582Jang,1997AJ....114..565Jang}; \citealt*{1999A&AS..135..477Romero}; 
\citealt{2002A&A...390..431Romero}; \citealt*{1993MNRAS.262..963GK, 1993A&A...271...89GK, 1995MNRAS.274..701GK}; 
\citealt{2000MNRAS.314..815GK,2003ApJ...586L..25GK,2011MNRAS.416..101GK};
\citealt*{1996MNRAS.281.1267Sagar}; \citealt{2004MNRAS.348..176Sagar};
\citealt{1991AJ....101.1196Carini, 1992AJ....104...15Carini, 2007AJ....133..303Carini}; 
\citealt*{1990AJ....100..347Carini, 1992ApJ...385..146Carini, 1998AJ....116.2667Carini, 2003AJ....125.1811Carini}; 
\citealt{2004MNRAS.350..175Stalin, 2004JApA...25....1Stalin, 2005MNRAS.356..607Stalin};
\citealt{1997AJ....113.1995Noble};
\citealt{2007BASI...35..141AG, 2009MNRAS.399.1622AG, 2010MNRAS.401.2622AG, 2012A&A...544A..37AG};
\citealt*{2005A&A...440..855Gupta, 2009NewA...14...88Gupta};
\citealt{1998ApJ...501...69Diego}; \citealt{2009AJ....138..991Ramirez}; \citealt{2011MNRAS.412.2717Joshi};
\citealt{2008AJ....136.2359Gupta, 2008AJ....135.1384Gupta, 2012MNRAS.425.1357Gupta};
\citealt{2010ApJ...719L.153Rani, 2010MNRAS.404.1992Rani, 2011MNRAS.413.2157Rani};
\citealt{2010ApJ...718..279Gaur, 2012MNRAS.425.3002Gaur}).
Since 1990, most observations of intra-night optical
variability (INOV) have been made using CCD detectors, which allow
simultaneous recording of a number of stars within the same chip.
Not only are some of these simultaneously monitored stars used
for measuring any variations in the seeing disk during the course
of the monitoring session, but, more importantly, they are used
as non-varying standards relative to which the light curve of the
target AGN can be drawn.  Such `differential light curves' (DLCs) are 
also drawn for the candidate `comparison stars' themselves and used 
to check for the presence of INOV of those stars, in which case they 
are disqualified as comparison stars (e.g., 
\citealt{1991Sci...254R1238Miller}; \citealt{2004JApA...25....1Stalin, 
2006ASPC..350..183Wiita}).
A key advantage of using DLCs is that the effects of any
fluctuations in the atmospheric attenuation and even in the seeing
disk are mostly canceled out, and this way the
variability detection threshold is pushed down enormously (e.g.,
\citealt*{1986PASP...98..802Howell}; \citealt*{1989Natur.337..627Miller}; 
\citealt{1993AJ....106.2441Gilliland};
\citealt{2005PASP..117.1187Howell}). Thus, intra-night optical
variability (INOV) with amplitudes as low as 1 to 2 per cent can be
routinely detected using 1-metre class telescopes.
(e.g., see \citealt{2012A&A...544A..37AG} and references therein).
Since 1998, a large body of such sensitive observations has been
accumulated, in a fairly uniform manner, using the 104-cm
Sampurnanand telescope of ARIES in Nainital (India) 
(\citealt{2004MNRAS.350..175Stalin, 2004JApA...25....1Stalin, 2005MNRAS.356..607Stalin};
\citealt{2008AJ....136.2359Gupta, 2008AJ....135.1384Gupta, 2012MNRAS.425.1357Gupta};
\citealt{2003ApJ...586L..25GK,2011MNRAS.416..101GK}; 
\citealt{2007BASI...35..141AG, 2009MNRAS.399.1622AG, 2010MNRAS.401.2622AG, 2012A&A...544A..37AG}
). Usually, the
targets monitored in these studies are optically luminous and relatively bright
point-like AGN, namely, quasars (both radio-loud and radio-quiet)
and BL Lacs, in the magnitude range $m_v$ = 15 - 17 mag.
\\
A number of statistical tests have been employed in the
literature for detecting the presence of variability in DLCs.
Until recently, the most popular test has been the, so called,
$C$-test (\citealt*{1997AJ....114..565Jang}; 
\citealt{1999A&AS..135..477Romero}). 
Basically, this involves computation of a factor `C' for a given DLC
of a target object, 
where C is the ratio of the standard deviation of the AGN light 
curve to the standard deviation of the comparison star-star light curve, i.e.,
\begin{equation}
C = \frac{\sigma_{t-s}}{\sigma_{s-s}} = 
\frac{\sigma_{t-s}}{\langle \sigma_{t-s} \rangle} 
\end{equation}
where $\sigma_{t-s}$ is the standard deviation of the `target-star' DLC,
and $\langle \sigma_{t-s} \rangle$ is the mean of the (formal) rms errors 
of the individual data points in the `target-star' DLC.
\\
This ratio `C' has been taken to have a Gaussian (normal) distribution (e.g., 
\citealt*{1997AJ....114..565Jang}, \citealt{1999A&AS..135..477Romero}). 
Thus, an AGN DLC found to have `C' greater than 
2.576 (corresponding to significance level, $\alpha=$ 0.01) 
is declared to be `variable'. Similarly, an AGN 
DLC having computed `C' value greater than 1.950 and less than 2.576
(corresponding to $\alpha=$ 0.05) is termed as `probable variable'.
However, recently, \citet{2010AJ....139.1269Diego} has questioned the validity
of this test on the ground that C-statistics does not have
a normal distribution and the two tailed p-values of 
normal distribution should not be 
used as a statistical indicator of INOV at a given $\alpha$ 
(variable vs. non-variable). The argument is as follows :\\ 
(a) The C-statistic is always positive, making it a 
{\it one-sided} comparison, unlike the normal 
Gaussian distribution which is {\it two-sided} comparison. \\
(b) For a test statistic to have a standard 
normal distribution, the expected value is distributed around 
0 while in case of `C' statistic it is distributed around 1 
when $\sigma_{t-s} = \sigma_{s-s}$ is satisfied.  \\
(c) One cannot compare two standard deviations using the normal 
distribution as they are not lineal statistical operators.   
\\
Thus, \citet{2010AJ....139.1269Diego} has argued in favour of {\it F}-test 
which relies on the computation of {\it F}-factor, being the ratio of 
two variances, as follows (see also, \citealt*{2010ApJ...723..737Villforth}):
\\
\begin{equation}
\label{ftest}
F =\frac{Var_{observed}}{Var_{expected}}= \frac{Var_{t-s}}{Var_{s-s}} 
=\frac{Var_{t-s}}{\langle \sigma_{t-s}^2 \rangle}
\end{equation}
\\
where $Var_{t-s}$ is the variance of the `target-star' DLC, and
$\langle \sigma_{t-s}^2 \rangle$,
is the mean of the squares of the (formal) rms errors of the individual data
points in the `target-star' DLC.
\\
\\
Clearly, both the {\it C}-test and the {\it F}-test require a precise estimate 
of the rms error ($\sigma$) associated with individual data points, 
which is usually determined using the {\sc APPHOT} routine in the 
{\emph IRAF}\footnote{\textrm 
{Image Reduction and Analysis Facility (http://iraf.noao.edu/) }} software. 
Many years ago, it was pointed out that the
$\sigma$ returned by this algorithm is systematically too low by
a factor, $\eta$, for which a value of 1.75 was estimated using
the DLCs derived for pairs of steady stars (\citealt{1995MNRAS.274..701GK}). 
This inference ($\eta \neq $ 1) has been borne out in several
independent studies from atleast 4 different observatories 
and the derived values of this parameter range
between 1.3 and 1.75 (\citealt{1995MNRAS.274..701GK}; \citealt{1999MNRAS.309..803Garcia}; 
\citealt{2005MNRAS.358..774Bachev}; \citealt{2004JApA...25....1Stalin}; 
\citealt{2007BASI...35..141AG}).
The most recent attempt to determine $\eta$ used DLCs for 73 pairs of steady 
stars and a best-fit value of $\eta$ = 1.5 was obtained
(\citealt{2012A&A...544A..37AG}). 
Clearly, a neglect of $\eta$ factor (i.e., setting
$\eta$ = 1) might often lead to spurious claims of INOV (above
a preset statistical significance threshold). It is therefore
important to achieve a greater precision in the determination of $\eta$,
by avoiding the use of any photometric data that fall within a
parameter space that is more prone to introducing larger uncertainty in
the $\eta$ determination.
\\
A prime candidate for a part of this `undesirable' parameter space is the
mismatch between the brightness of the chosen steady comparison  
stars which are paired to derive the DLCs which are collectively 
used for $\eta$ determination.
The mismatch can be represented by $\Delta m_{s}$ = $m_{s1} - m_{s2}$.
The purpose of the present study is to identify the `safe' 
parameter space for $\Delta m_{s}$, outside which 
a significant distortion of the
$\eta$ estimate can occur. This has important implications for the INOV 
search since several claims of large INOV of AGN have been questioned 
because of a large mismatches between their brightnesses and those of the 
comparison stars used for deriving the differential light curves (e.g., \citealt{2007MNRAS.374..357Cellone}).  
\\
\section{The sample of intra-night optical DLCs}
\label{sample}
Using the 1-m Sampurnanand telescope (ST) of ARIES, a long-term programme 
was launched in 1998, for characterizing the INOV properties of important 
AGN classes.
Results of this ongoing study have been reported in a 
series of publications and in the Ph.D. theses of C. S. Stalin (\citeyear{2003PhDT.......263Stalin}) 
and Arti Goyal (\citeyear{2010PhDT.......263Arti}) (\citealt{2012A&A...544A..37AG} and references 
therein; \citealt{2005MNRAS.356..607Stalin} and references therein). 
Optical intra-night monitoring data from other optical observatories 
in India, such as the 2-m Himalayan Chandra Telescope
(HCT) and the 2.4m Vainu Bappu Telescope (VBT) of IIA, the 1.2m telescope 
at the Gurushikhar observatory of PRL and the 2-m IUCAA Girawali 
Observatory (IGO) telescope of IUCAA were also obtained to augment the 
data taken with the 1-m ST. Nearly always, just one target AGN was monitored 
on a given night.  
\\
The above intra-night monitoring program has covered 22 radio-quiet 
quasars (RQQs), 10 radio-intermediate quasars (RIQs), 
9 radio lobe-dominated quasars (LDQs), 11 radio core-dominated 
quasars showing high optical polarization (HPCDQs) 
and 12 showing low optical polarization (LPCDQs), as well 
as 13 TeV detected BL Lac objects. Sources in the various 
classes were chosen from the catalog of 
\citet{2001yCat.7224....0Veron} and its subsequent releases.  
All the sources lie at $z$ $> 0.14$ and have a listed 
$m_B < 18$mag, which allows enough signal-to-noise ratio (SNR) in a typical 
exposure time of $\sim10$ minutes. 
Each source was monitored for a minimum duration of $\sim$4 hours. 
These CCD monitoring observations, aided by a careful and uniform
data analysis procedure, have routinely allowed INOV detection with 
amplitude ($\psi$) as low as 1 - 2 per cent. 
The present sample consists of 262 such intra-night observations 
obtained from the entire data set from our ARIES AGN INOV programme. 
\\
\section{Observations and data analysis}
\label{obs}
\begin{figure}
\includegraphics[height=  10.0cm, width = 12.0cm]{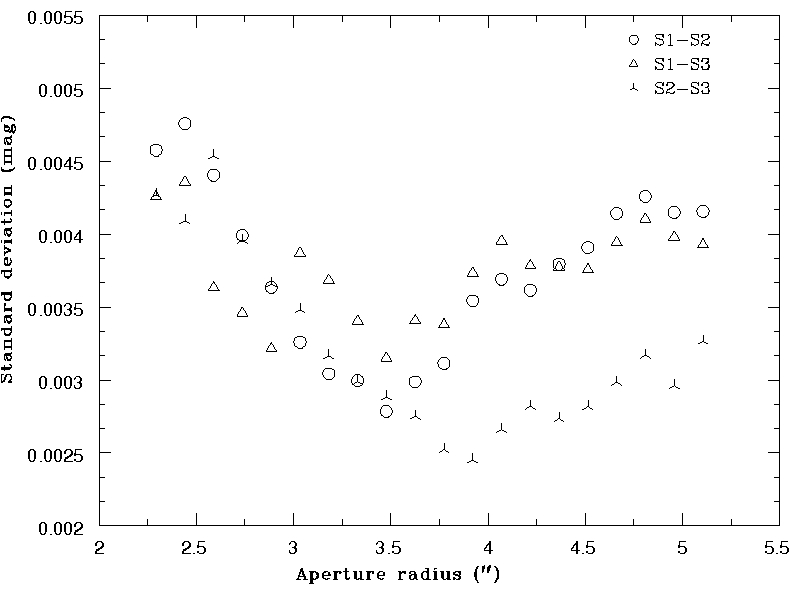}
\caption{ 
The rms of the DLCs derived for a pair of (steady) comparison stars used 
for the target quasar J2203$+$3145, versus photometric aperture radius, 
monitored on 15-Sep-2007. The minimum in standard deviation on that 
night is seen to occur for an aperture radius $\simeq$ 3.8 arcsec. }
\label{aper_rad}
\end{figure}

The observations were made mostly in the {\it R} filter and
occasionally in the {\it V} filter.
The exposure time was typically between 10 to 20 minutes for  
the ARIES and Gurushikar observations and ranged between 3 to 6 minutes for 
the observations from VBT, IAO and IGO, depending on the
brightness of the source, the phase of the moon and the sky 
transparency on that night. The field positioning was adjusted 
so as to also have within the CCD frame at least 2--3 comparison stars.
For all the telescopes, bias frames were taken intermittently, 
and twilight sky flats were also obtained. 
\\
The pre-processing of the images (bias subtraction, flat-fielding and 
cosmic-ray removal) was done by applying the standard procedures in the 
{\textrm IRAF} and {\textrm MIDAS}\footnote{\textrm {Munich Image and 
Data Analysis System (http://www.eso.org/sci/data-processing/software/esomidas/) }}
software packages. The instrumental magnitudes 
of the target AGN (all point-like) and the stars in
the image frames were determined by aperture photometry, using
{\textrm APPHOT}. 
The magnitude of the target AGN was measured relative to 
a few apparently steady comparison
stars present on the same CCD frame. In this way 
DLCs for each AGN were derived 
relative to 2-3 comparison stars designated as S1, S2, S3.
\\
These comparison stars mostly lie within about 1.5 magnitude
of the target AGN, this being an important criterion for minimizing 
the possibility of spurious INOV detection 
(e.g., \citealt{2007MNRAS.374..357Cellone}). 
Spurious variability 
on account of different second-order
extinction coefficients for the AGN and their comparison 
stars is a possible problem if the colours of the objects are different.
Although the {\it B-R} colors of the AGN and the comparison 
stars used in our study often differ
significantly, it was shown by \citet{1992AJ....104...15Carini}  
and \citet{2004JApA...25....1Stalin} that  even though their 
photons travel through varying airmass
during the course of monitoring, this has a negligible effect on DLCs.
For each night, an optimum aperture radius for photometry
was chosen by minimizing the dispersions in the star-star DLCs, 
that were found using different aperture radii, starting 
from the median seeing (FWHM) value on that night to 4 
times that value (Fig. \ref{aper_rad}). For very small 
aperture radii, the scatter will be large due to improper photon 
counting statistics, as the total photon count from the source will 
be small. On the other hand, at very large aperture radii, 
the scatter will increase as the on-source measurement will be affected by the  
emission from the sky background (\citealt*{1989PASP..101..616Howell}). 
At intermediate aperture radii, a minimum will occur as shown in Fig. \ref{aper_rad}.
We selected the appropriate aperture for each night
as the one that provided the minimum dispersion for the DLC found among all pairs of 
the comparison stars, as the same stars would be used to produce DLCs  
against the target quasars to check for their INOV. Thus, using the aperture which 
provides minimum dispersion will set a threshold for INOV detection 
on that night (e.g., \citealt{2004JApA...25....1Stalin}). 
Typically, the selected aperture radius was 
$\sim$4$^{\prime\prime}$ and the seeing was $\sim$2$^{\prime\prime}$.
\\
\\
\\
\\
\section{Determination of $\eta$}
\label{eta_comp}
As mentioned in Sect. \ref{intro}, the photometric errors 
returned by {\sc APPHOT} are significantly underestimated. 
In this work, we make a fresh attempt to determine 
$\eta$ using our enlarged dataset of 262 DLCs from our
ARIES AGN monitoring program 
(see \citealt{2012A&A...544A..37AG}; Sect.\ \ref{sample}).  
Out of the 3 star-star DLCs available for each night 
(using the 3 comparison stars monitored), we first 
selected the steadiest (one having minimum variance) star$-$star DLC. Thus, for
our entire dataset we have got 262 `steady' DLCs, whose 524 stars 
appear to have not varied on the corresponding 
nights. For each selected DLC, with $N_p$ points, we then computed  
$\chi^2$ corresponding to its degree of freedom, $\nu$ =
$N_p - 1$, which is given as;

\begin{equation}
\label{chi2}
\chi^2 = \sum\limits_{i=1}^{N_p-1} \frac{1}{\sigma_i^2} (\Delta m_i - \langle \Delta m \rangle)^2
\end{equation}
where the expected value $\langle \Delta m \rangle$ is the sample mean of the DLC. 
$N_p$ is the number of data points in the lightcurve, $\Delta m_i$ is the
differential magnitude of the $i^{th}$ data point in the lightcurve
and $\sigma_i$
is the rms measurement error associated with each $\Delta m_i$.

To compute $\eta$, we use

\begin{equation}
\label{eta_equ}
\nu = \sum\limits_{i=1}^{N_p-1} \frac{1}{\eta^2 \sigma_i^2} (\Delta m_i - \langle \Delta m \rangle)^2
\end{equation}
where the degree of freedom $\nu$ is also the expected $\langle \chi^2 \rangle$ value 
for a pair of non-variable stars. The simplest approach is to use 
regression analysis given by

\begin{equation}
\label{regression}
\chi^2 = \eta^2 \nu + \epsilon
\end{equation}

where $\epsilon$ is the residual associated with each pair of $\chi^2$ and $\nu$.
However, we do not know that residuals are Gaussian distributed, 
or are homogeneous with respect to the values of independent variable, 
precluding a reliable least square fitting. As our regression analysis 
exhibit an ``expected value - residual'' we can transform 
the variables to stabilize the variance. The most common method is 
the Box-Cox set of tranformations (\citealt*{1964AJ....113.1995Box}; \citealt*{2005drea.book.....Box}).
In our case this involves using logrithms of the $\chi^2$ values 
to homogenize the variance of regression analysis and to maintain 
the linear relationship
between the $\chi^2$ and $\nu$, we tranform $\nu$ to $log(\nu$). Then, we 
fix the slope to 1 in the regression analysis to obtain :
  
\begin{equation}
\label{log_regress}
log(\langle \chi^2 \rangle) = K + log(\nu)
\end{equation}

where $\eta^2 = 10^K$. The error in $\eta^2$ is computed using \citet*{2003drea.book.....Bevington}

\begin{equation}
\label{error_eta}
\sigma_\eta^2 = \eta^2 \times (2.303 \times \sigma_K)^2
\end{equation}
where $\sigma_K$ is the error in $K$.
Using these, we obtain $\eta = 1.54 \pm 0.05$ for the 
entire set of 262 steady `star-star' DLCs data listed in Table \ref{result}.   

\begin{figure}
\hspace*{0.0cm}
\includegraphics[height=9.0cm,width=11.0cm]{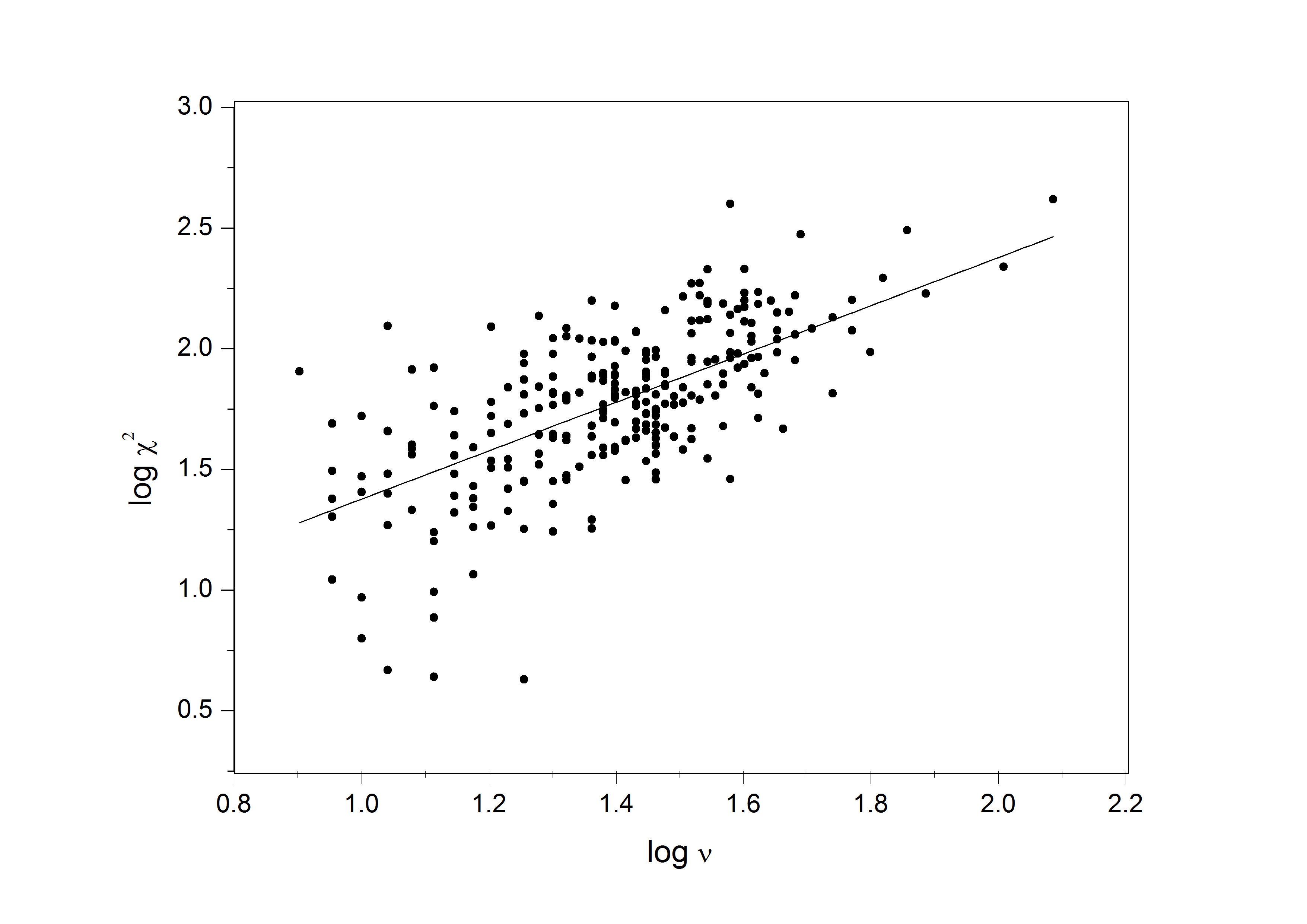}
\caption{
Plot of $\chi^2$ values against degrees of freedom, computed for our entire data 
set of 262 night. The solid line gives the slope fixed at 1
(see sect. \ref{eta_comp}). }
\label{etafig}
\end{figure}

\begin{figure}
\hspace*{0.0cm}
\includegraphics[height=5.0cm,width=7.0cm]{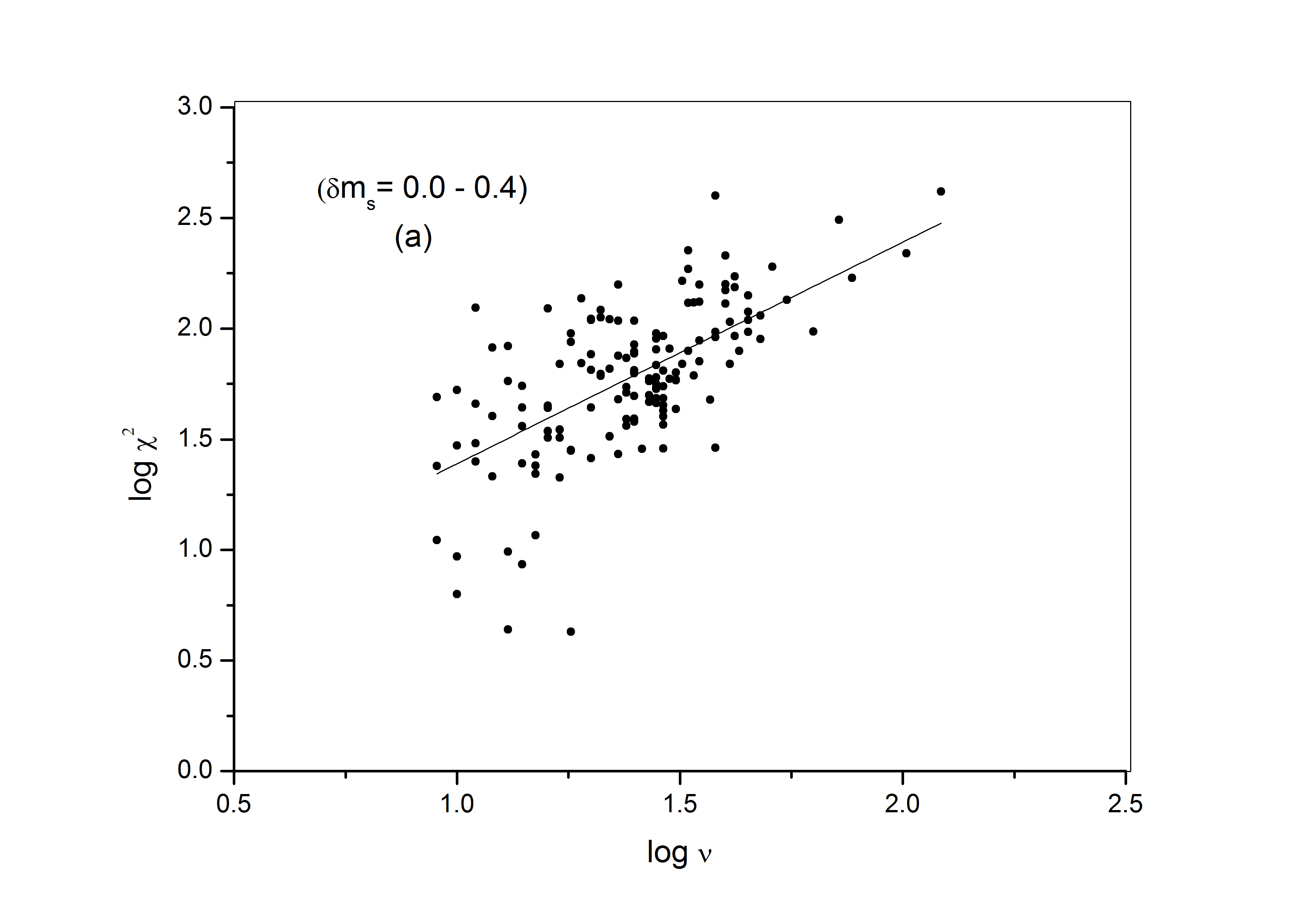}
\includegraphics[height=5.0cm,width=7.0cm]{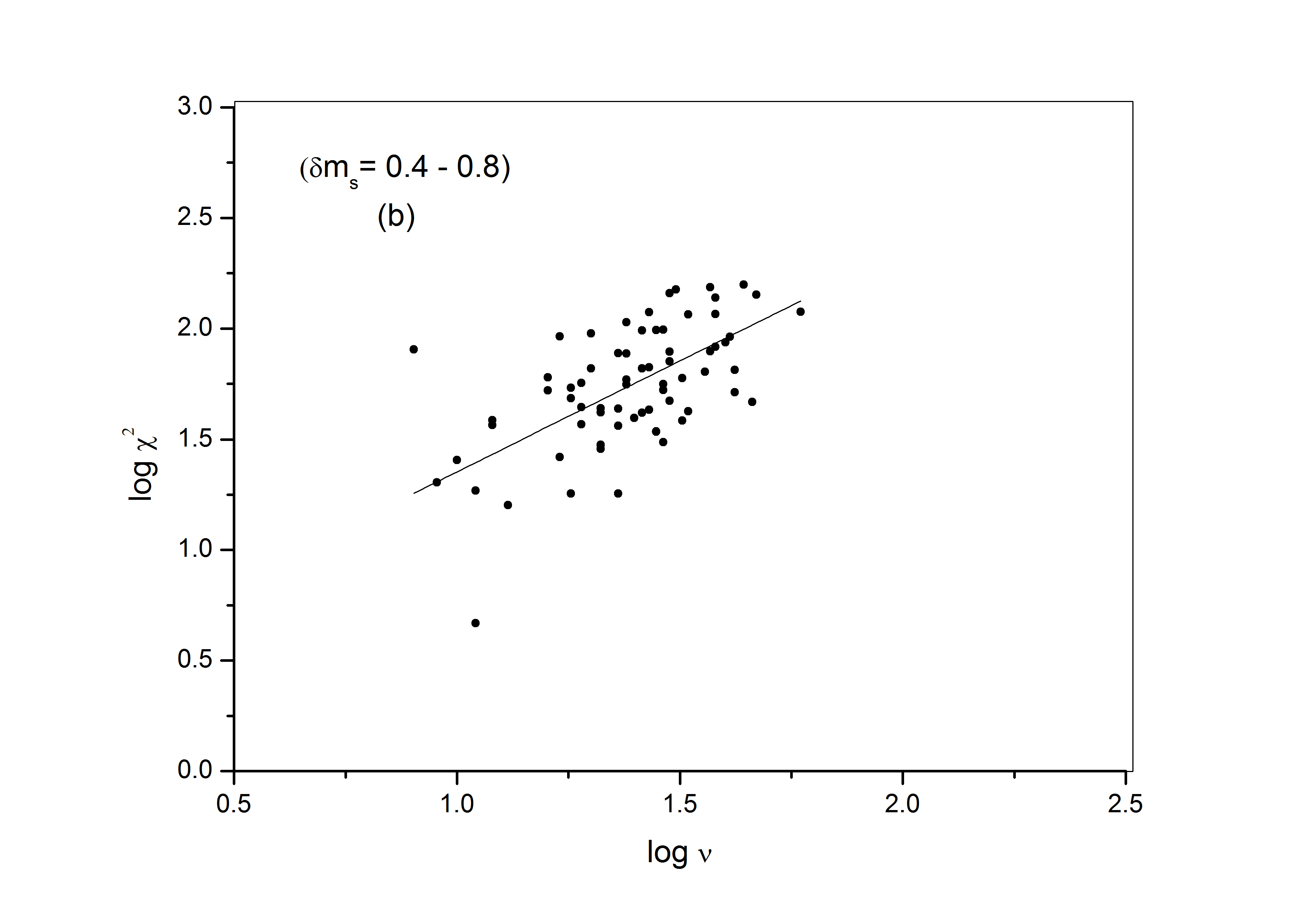}
\includegraphics[height=5.0cm,width=7.0cm]{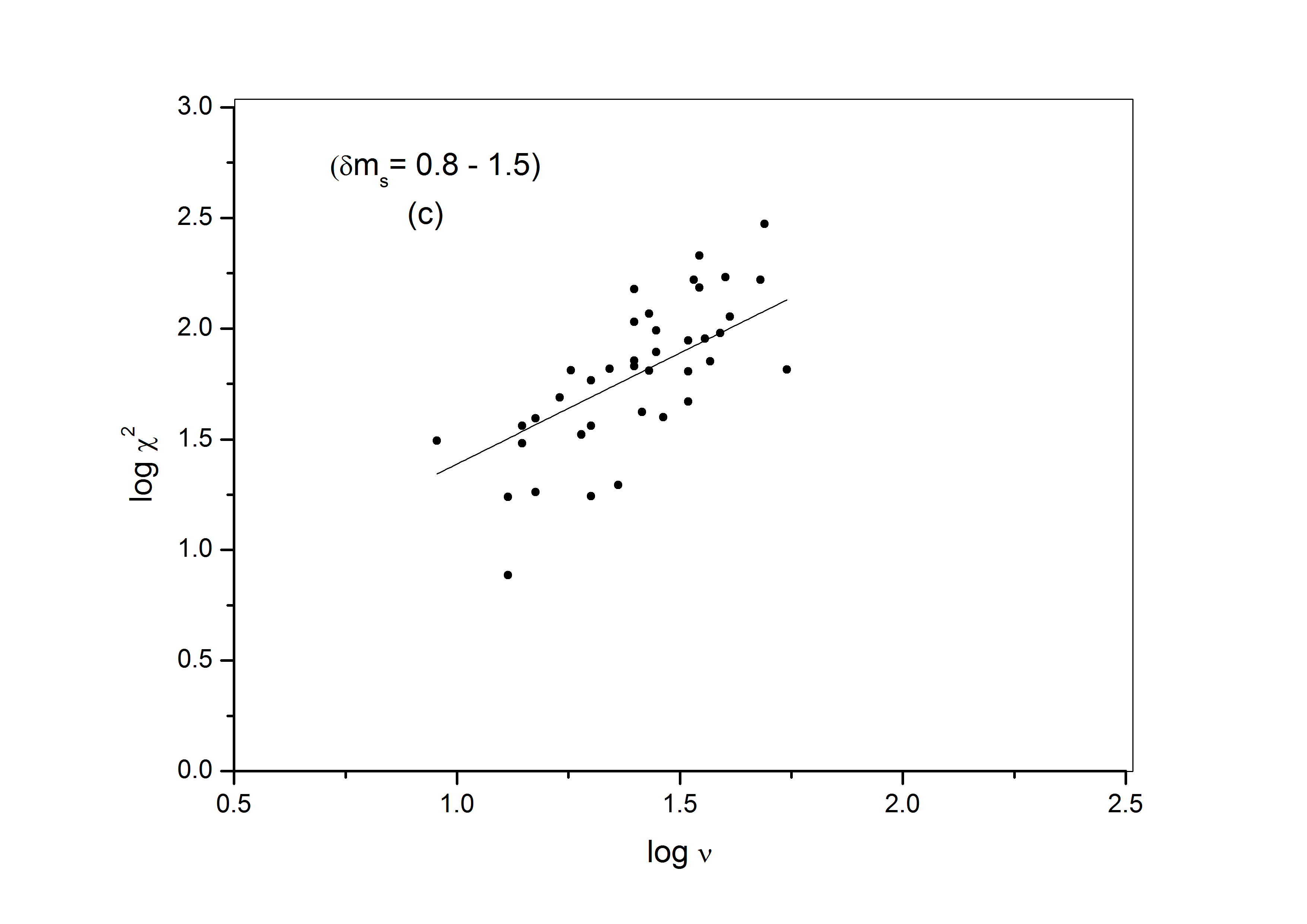}
\caption{
Plot of $\chi^2$ values against degrees of freedom, 
computed for the 3 ranges of apparent magnitude 
difference between the (steady) stars paired to derive the DLCs. (a) 
$\chi^2$ for the $\Delta m_{s} = 0.00 - 0.40$ (148 DLCs); 
(b) $\chi^2$ for the $\Delta m_{s} = 0.40 - 0.80$ (69 DLCs) and 
(c) $\chi^2$ for the $\Delta m_{s} = 0.80 - 1.50$ (39 DLCs).
The solid line shows slope of regression analysis fixed at 1
(see Sect.\ \ref{eta_comp}). }
\label{etamag}
\end{figure}

In Fig. \ref{etafig}, we plot for all 262 `steady' star-star DLCs, 
the computed $\chi^{2}$ values against the respective values of $\nu$.
Accodingly, we adopt $\eta=$1.54, for scaling up the $\emph IRAF$
photometric rms errors (see Sect.\ 5).
\\
\\
\\
As mentioned in Sect.\ref{intro}, a principal goal of the present
study is to check the dependence of $\eta$ on the brightness 
mismatch between the stars which are paired to derive the 
`steady' star-star DLCs. For this, we divide our sample of 
262 DLCs into  subsamples corresponding to three intervals of 
the apparent magnitude difference ($\Delta m_s$) between the star-pair
(see column 7 of Table \ref{result}). 
These subsamples have $\Delta m_s$ in the ranges 0.00-0.40 mag (148 DLCs), 
0.40-0.80 mag (69 DLCs) and 0.80 to 1.50 mag (39 DLCs). 
Out of the 262 DLCs star-star DLCs considered here, only 6 have $\Delta m_s > 1.50$ mag. 
The computed values of $\chi^2$ for the three subsamples are 
plotted in Fig. \ref{etamag}. We apply the regression 
analysis, as explained above, to compute the $\eta$ values 
for these subsamples. These values of $\eta$ are found to be 
1.56 $\pm$ 0.07, 1.50 $\pm$ 0.09 and 1.56 $\pm$ 0.13 
for the subsamples defined by 
0.00 $< \Delta m_s < $0.40, 0.40 $< \Delta m_s < $0.80
and 0.80 $< \Delta m_s < $1.50, respectively.   
We note that these values of $\eta$ are mutually consistent 
for the three magnitude bins.
We thus conclude that the determination of $\eta$ is essentially independent of the brightness 
mismatch of at least up to 1.5 mag between the comparison stars used.
\\
\\
\\
\section{Discussion}

In order to counter-check these findings, we now subject our analysis 
to a {\it sanity check} (Table \ref{result}). 
For this we have computed the expected number of {\it false 
positives} ({\it `Type 1 error'}) for our dataset of 262 DLCs. We have 
thus performed the $F-test$ (Eq.\ \ref{ftest}) on the 262 steady star-star 
DLCs after accounting for the photometric error underestimation factor (i.e., 
replacing the denominator with $\eta^2 \sigma^2$ in Eq.\ \ref{ftest}). 
The expression for $F$ is given by $F_{\nu_1,\nu_2}^{\alpha} = \sigma_1^2/\sigma_2^2 $ 
where $\sigma_1$ and $\sigma_2$ are the variances of the numerator and
the denominator and the $\nu_1$ and $\nu_2$ are the corresponding 
degrees of freedom. 
In our analysis, we have simplified the $F$ expression to $F_{\nu}^{\alpha}$ as
$\nu_1$ = $\nu_2 = \nu$ is the degree of freedom for the `star-star' DLC.
In this way, the  $F$-value was computed for each DLC and compared with 
the critical $F$-value. Recall that smaller the $\alpha$, 
the less likely is it to occur by chance. For the present study, we 
have used two values of significance level, $\alpha$ = 0.01 and 0.05.
Thus we claim a spurious INOV detection for a DLC, 
when the computed $F-$ value exceeds
the critical $F-$value at $\alpha = $ 0.01. We thus assign a `variable'
designation (V) to it. We assign a `probable variable' (PV) designation when 
the computed $F-$ value is found to be between the 
critical $F-$ values at $\alpha = $ 0.01 and 0.05, otherwise 
`non-variable' (N) designation as assigned to the `star-star' DLC.
\\
Following this analysis, out of 262 steady `star-star' DLCs, 
6 DLCs were found to be of `V' type, while 12 were designated 
as `PV' (Table \ref{result}).  At $\alpha = 0.01$ (i.e., $p>$  0.99), 
we expect among the 262 `star-star' DLCs, $\sim$3 DLCs to 
be falsely classified as `V'. Similarly, at $\alpha = 0.05$ 
(i.e., $p>$ 0.95), the expected number of false positives 
is $\sim$13. We find that for our analysis, the {\it observed} 
number of false positive is 6 at $\alpha =$ 0.01 and 18 at 0.05. 
Since the distribution of false positives (Type 1 errors)
is binomial, we expect its actual number for a given test will be between 0  
and 9 and in most cases between 3$\pm$2 at $\alpha = 0.01$. Similarly, at $\alpha = 0.05$,
the actual number of false positives will be between 2 and 24 and 
in most cases will be 13$\pm$ 4. The good match between the {\it observed} 
and {\it expected} values of {\it false positives} validates our 
analysis procedure adopting $\eta$ = 1.54 as determined here.    
\\

Also, for our three subsamples defined in Sect. 4, we find the 
expected numbers of false postives for most cases will be 
2$\pm$1 (148 DLCs in the magnitude bin $ 0.0 < \Delta m_s < 0.4 $),
1$\pm$1 (69 DLCs in the magnitude bin $ 0.4 < \Delta m_s < 0.8 $) and
1$\pm$1 (39 DLCs in the magnitude bin $ 0.8 < \Delta m_s < 1.5 $) at $\alpha = 0.01$.
We find that the {\it observed} numbers for false positives are 4, 1 and 1.
Similarly, at $\alpha = 0.05$, expected numbers of false postives for most cases will be
7$\pm$3 (148 DLCs in the magnitude bin $ 0.0 < \Delta m_s < 0.4 $),
4$\pm$2 (69 DLCs in the magnitude bin $ 0.4 < \Delta m_s < 0.8 $) and
2$\pm$2 (39 DLCs in the magnitude bin $ 0.8 < \Delta m_s < 1.5 $) at.
We find that the {\it observed} numbers for false positives are 10, 3 and 5, respectively.   
This again shows a close match between the observed 
and expected values of {\it false positives}, validating the 
estimate of $\eta = 1.54$ up to a magnitude mismatch of $\sim$1.5
mag between the comparison star pairs. 

The vast majority of the data analyzed here comes from ST and therefore
our results strictly apply to those observations. The data from the HCT, IGO, GSO 
and VBT all seem consistent with the ST results, but each of these 
telescopes contributed measurements that are not numerous enough 
to perform useful separate analyses for these telescopes.
Therefore we cannot yet determine whether
the value of $\eta$ we have found is a fundamental feature of $\textrm IRAF$'s 
$\textrm APPHOT$ and thus universal, or somewhat dependent on 
the telescope and the instrument used. Over the next couple of years we
anticipate obtaining comparably large data sets with a new ARIES 1.3 m
telescope located at a different site near Nainital. We will perform a similar
analysis of the values of $\eta$ for those additional data and that will 
lead us to a better grasp of the root of this error underestimation.
We do, however, note that because the seeing varied substantially 
(from 0.7 to 3.5 arc sec) for the data we have employed here, the value
of $\eta$ does seem to be fairly independent of this important aspect of 
the differential photometry process.

\section{Summary}
In this study, we have determined the photometric error underestimation factor 
$\eta$ applicable to point-source aperture photometry carried out
using the {\emph IRAF} (APPHOT) software. For this we have used an 
unprecedentedly large set of 262 DLCs taken on 262 nights, about 85 per cent 
of which are taken with the 1-m telescope (ST) of ARIES. By subjecting 
this large database to 
a $\chi^2$ analysis we 
find that $\eta$ = 1.54$\pm$0.05, which is consistent with the 
most recently published estimate of this important parameter, 
which was derived using a $\sim$4 times smaller sample of DLCs than we 
have used here 
(see, \citealt{2012A&A...544A..37AG}). A sanity check, based on 
the computation of `false positives' employing the $F$-test, 
was performed and it has validated the estimate of $\eta$ = 1.54.
\\
We have further checked for any dependence of the $\eta$ factor on the 
apparent magnitude mismatch ($\Delta m_{s}$) between the comparison stars 
paired (taking them to be steady, as inferred from inspection of their DLCs). 
For this we divided our sample of DLCs into three subsamples, characterized 
by $0.0 < \Delta m_s < 0.40$ (148 DLCs),  $0.40 < \Delta m_s < 0.80$ (69) 
and $0.80 < \Delta m_s < 1.50$ (39 DLCs). 
For each subsample the sanity check again showed consistency with $\eta$ = 1.54.
It is thus concluded that $\eta$ = 1.54
remains valid even when the magnitudes of the `steady' stars paired to 
derive a DLC differ by as much as 1.5 mag. In other words, even a magnitude
difference of up to 1.5-mag between the two stars paired to derive a DLC and $\eta$, 
should not result in a spurious claim of INOV for either of the two stars. 
As a corollary, it can be reasonably asserted that deriving DLCs of (point-like) AGN 
using a comparison star that is within about 1.5 magnitude of the AGN, should 
not lead to spurious claim of INOV for the AGN. However, this could well 
be the case for significantly larger magnitude mismatches, as argued
by \citet{2007MNRAS.374..357Cellone} in the context of some claims of 
dramatic INOV.

The present analysis is dominated by the $R-$band data taken using the 
ARIES 1-m telescope (ST). Therefore, the present conclusion strictly 
apply only to the $R-$band taken with this telescope.
In the coming years, we plan to expand the present analysis to observation 
taken with the 1.3-m Devasthal Optical Telescope (DOT) recently installed
at a site well removed from that of the ST.   
\\
\begin{deluxetable}{ccccccccccccc}
\tablecolumns{13}
\tabletypesize{\tiny}
\tablecaption{Summary of observations and derived variability status for the `steady' star-star DLCs \label{result}}
\tablewidth{0pt}
\tablehead{
AGN  & Obs. date    & Tel.{$^\P$}  &  Filter.  & Dur. & $N_p$ & $\Delta m_{s}$   & $\sigma $  &  std dev. & $\chi^2_{s}$   &  $F_{s}$ &  Status$^\dag$ & Ref$^\pounds$.  \\
name    &  dd.mm.yy& used      & used     &   (hr)  &      &  (mag) & ($10^{-2}$ mag)  & ($10^{-2}$ mag)     &    &           &         &                       \\
        (1) & (2) & (3) & (4) & (5) & (6) & (7) & (8) & (9)  & (10) & (11) & (12) & (13) \\ }
\startdata
\multicolumn{13}{l}{\small Radio quiet quasars (RQQs)[22 sources; 68 DLCs]}\\ \hline
&&&&&&&&&&&&\\
J0045$+$0410 & 21.10.98 & ST  & R & 2.39 & 14 & 0.376 & 0.5 & 0.2  &  4.37  & 0.13 &N  &  (a) \\ 
J0045$+$0410 & 05.11.98 & ST  & R & 3.21 & 30 & 0.369 & 0.8 & 1.0  & 40.05  & 0.55 &N  &   (a)\\  
J0045$+$0410 & 16.10.04 & HCT & R & 6.04 & 25 & 1.859 & 0.1 & 0.2  & 79.36  & 1.24 &N  &   (b)\\
&&&&&&&&&&&&\\
J0103$+$0321 & 05.11.05 & HCT & R & 5.94 & 21 & 1.093 & 0.3 & 0.4  & 36.32  & 0.72 &N  &   (b)\\
J0103$+$0321 & 05.11.05 & ST  & R & 5.83 & 20 & 0.570 & 0.4 & 0.6  & 44.09  & 1.02 &N  &   (b)\\
&&&&&&&&&&&&\\
J0239$-$0001 & 06.11.05 & HCT & R & 6.42 & 19 & 0.779 & 0.1 & 0.2  & 53.90  & 1.26 &N  &   (b)\\
&&&&&&&&&&&&\\
J0516$-$0027 & 10.12.01 & ST  & R & 5.77 & 23 & 0.160 & 0.3 & 0.3  & 32.50  & 0.56 &N  &   (c)\\
J0516$-$0027 & 19.12.01 & ST  & R & 7.52 & 35 & 0.210 & 0.3 & 0.5  & 131.13 & 1.16 &N  &   (c)\\
J0516$-$0027 & 20.11.03 & HCT & R & 7.28 & 39 & 0.264 & 0.1 & 0.2  & 96.37  & 1.07 &N  &   (b)\\
J0516$-$0027 & 18.11.04 & ST  & R & 6.29 & 34 & 0.282 & 0.1 & 0.2  & 79.19  & 1.01 &N  &   (b)\\
J0516$-$0027 & 16.12.04 & HCT & R & 6.79 & 34 & 1.256 & 0.2 & 0.2  & 63.96  & 0.60 &N  &   (b)\\
&&&&&&&&&&&&\\
J0751$+$2919 & 14.12.98 & ST  & R & 7.41 & 40 & 1.569 & 0.3 & 0.6 & 145.70  & 1.57 &N  &   (c)\\
J0751$+$2919 & 13.01.99 & ST  & R & 8.32 & 56 & 0.362 & 0.3 & 0.5  & 134.70 & 0.93 &N  &   (c)\\
J0751$+$2919 & 24.11.99 & ST  & R & 5.39 & 28 & 0.702 & 0.3 & 0.3  & 42.90  & 0.62 &N  &   (c)\\
J0751$+$2919 & 09.12.99 & ST  & R & 6.21 & 31 & 0.710 & 0.2 & 0.5  & 144.38 & 2.13 &PV &   (c)\\
J0751$+$2919 & 01.12.00 & ST  & R & 5.95 & 32 & 0.372 & 0.3 & 0.4  & 63.30  & 0.78 &N  &   (c)\\
J0751$+$2919 & 25.12.01 & ST  & R & 5.44 & 30 & 0.372 & 0.4 & 0.4  & 36.78  & 0.54 &N  &   (c)\\
J0751$+$2919 & 17.12.04 & HCT & V & 3.69 & 15 & 0.318 & 0.1 & 0.2  & 24.61  & 0.74 &N  &   (b)\\
J0751$+$2919 & 17.12.04 & ST  & R & 7.02 & 34 & 0.238 & 0.1 & 0.3  & 130.66 & 1.56 &N  &   (b)\\
J0751$+$2919 & 12.01.05 & ST  & R & 7.15 & 16 & 0.129 & 0.1 & 0.2  & 22.10  & 0.61 &N  &   (b)\\
J0751$+$2919 & 07.03.06 & HCT & R & 8.06 & 29 & 0.046 & 0.1 & 0.2  & 55.95  & 0.84 &N  &   (b)\\
J0751$+$2919 & 07.03.06 & ST  & R & 8.33 & 46 & 0.079 & 0.1 & 0.2  & 141.20 & 1.32 &N  &   (b)\\
&&&&&&&&&&&&\\
J0827$+$0942 & 27.12.98 & ST  & R & 8.15 & 60 & 0.415 & 0.3 & 0.4  & 119.04 & 0.88 &N  &   (a)\\
J0827$+$0942 & 13.01.05 & HCT & V & 6.47 & 16 & 0.061 & 0.1 & 0.2  & 24.02  & 0.67 &N  &   (b)\\
J0827$+$0942 & 13.01.05 & ST  & R & 6.94 & 17 & 0.000 & 0.1 & 0.2  & 44.74  & 1.18 &N  &   (b)\\
&&&&&&&&&&&&\\
J0835$+$2506 & 25.12.98 & ST  & R & 4.68 & 26 & 0.911 & 0.4 & 0.6  & 67.49  & 1.13 &N  &   (a)\\
J0835$+$2506 & 14.01.99 & ST  & R & 8.91 & 78 & 0.206 & 0.4 & 0.6  & 169.41 & 0.92 &N  &   (a)\\
J0835$+$2506 & 10.12.99 & ST  & R & 6.72 & 33 & 0.714 & 0.4 & 0.6  & 59.65  & 0.75 &N  &   (a)\\
&&&&&&&&&&&&\\
J0853$+$4349 & 17.02.99 & ST  & R & 7.70 & 39 & 0.234 & 0.4 & 0.7  & 91.50  & 0.99 &N  &   (a)\\
&&&&&&&&&&&&\\
J0935$+$4331 & 20.02.99 & ST  & R & 4.47 & 26 & 0.883 & 0.2 & 0.3  & 106.91 & 1.69 &N  &   (a)\\
&&&&&&&&&&&&\\
J0938$+$4128 & 27.03.99 & ST  & R & 2.73 & 17 & 0.000 & 0.5 & 0.6  & 34.40  & 0.73 &N  &   (a)\\
&&&&&&&&&&&&\\
J0948$+$4335 & 15.01.99 & ST  & R & 7.97 & 44 & 0.209 & 0.3 & 0.5  & 79.10  & 0.80 &N  &   (c)\\
J0948$+$4335 & 26.02.00 & ST  & R & 7.97 & 39 & 0.490 & 0.4 & 0.6  & 82.44  & 0.91 &N  &   (c)\\
J0948$+$4335 & 23.01.01 & ST  & R & 6.73 & 25 & 0.505 & 0.3 & 0.6  & 77.02  & 1.20 &N  &   (c)\\
&&&&&&&&&&&&\\
J1019$+$2744 & 14.03.99 & ST  & R & 7.32 & 43 & 0.304 & 0.5 & 0.7  & 92.57  & 0.86 &N  &   (c)\\
J1019$+$2744 & 14.01.00 & ST  & R & 7.08 & 34 & 0.441 & 0.2 & 0.2  & 42.28  & 0.52 &N  &   (c)\\
J1019$+$2744 & 27.02.00 & ST  & R & 8.81 & 37 & 0.442 & 0.2 & 0.3  & 63.77  & 0.61 &N  &   (c)\\
&&&&&&&&&&&&\\
J1032$+$3240 & 13.03.99 & VBT & V & 8.40 & 45 & 0.503 & 0.5 & 0.8  & 158.00 & 1.16 &N  &   (c)\\
J1032$+$3240 & 02.03.00 & ST  & R & 4.95 & 19 & 0.887 & 0.2 & 0.4  & 64.64  & 1.45 &N  &   (c)\\
J1032$+$3240 & 05.04.00 & ST  & R & 6.17 & 24 & 0.136 & 0.1 & 0.3  & 108.19 & 1.85 &PV &   (c)\\
J1032$+$3240 & 23.03.01 & ST  & R & 6.84 & 25 & 0.303 & 0.5 & 0.6  & 51.32  & 0.83 &N  &   (c)\\
J1032$+$3240 & 06.03.02 & ST  & R & 8.53 & 34 & 0.134 & 0.2 & 0.3  & 185.91 & 1.28 &N  &   (c)\\
J1032$+$3240 & 08.03.02 & ST  & R & 8.31 & 24 & 0.127 & 0.2 & 0.3  & 75.22  & 1.17 &N  &   (c)\\
&&&&&&&&&&&&\\
J1104$+$3141 & 12.03.99 & ST  & R & 8.80 & 43 & 0.551 & 0.6 & 0.7  & 51.55  & 0.48 &N  &   (c)\\
J1104$+$3141 & 14.04.00 & ST  & R & 5.61 & 22 & 0.035 & 0.3 & 0.5  & 62.25  & 1.01 &N  &   (c)\\
J1104$+$3141 & 21.04.01 & ST  & R & 6.40 & 27 & 0.032 & 0.5 & 0.5  & 28.59  & 0.41 &N  &   (c)\\
J1104$+$3141 & 22.04.01 & ST  & R & 5.58 & 24 & 0.037 & 0.5 & 0.5  & 27.08  & 0.43 &N  &   (c)\\
&&&&&&&&&&&&\\
J1119$+$2119 & 14.04.05 & ST  & R & 5.02 & 30 & 0.065 & 0.1 & 0.2  & 48.37  & 0.70 &N  &   (b)\\
J1119$+$2119 & 30.03.06 & ST  & R & 6.17 & 41 & 0.072 & 0.1 & 0.3  & 149.11 & 1.57 &N  &   (b)\\
J1119$+$2119 & 31.03.06 & ST  & R & 4.25 & 26 & 0.070 & 0.1 & 0.2  & 49.47  & 0.83 &N  &   (b)\\
&&&&&&&&&&&&\\
J1246$+$0224 & 13.04.05 & ST  & R & 5.51 & 10 & 0.046 & 0.1 & 0.3  & 48.90  & 2.01 &N  &   (b)\\
&&&&&&&&&&&&\\
J1255$+$0144 & 22.03.99 & ST  & R & 7.46 & 43 & 0.483 & 0.4 & 0.5  & 64.91  & 0.59 &N  &   (c)\\
J1255$+$0144 & 09.03.00 & ST  & R & 6.14 & 29 & 0.144 & 0.1 & 0.2  & 80.28  & 1.05 &N  &   (c)\\
J1255$+$0144 & 03.04.00 & ST  & R & 4.32 & 21 & 0.154 & 0.1 & 0.4  & 109.28 & 2.53 &V  &   (c)\\
J1255$+$0144 & 26.04.01 & ST  & R & 4.60 & 20 & 0.107 & 0.2 & 0.5  & 136.56 & 1.88 &N  &   (c)\\
J1255$+$0144 & 18.03.02 & ST  & R & 7.88 & 25 & 0.130 & 0.4 & 0.3  & 73.50  & 0.36 &N  &   (c)\\
&&&&&&&&&&&&\\
J1424$+$4214 & 03.04.99 & ST  & R & 7.22 & 41 & 0.056 & 0.3 & 0.6  & 158.64 & 1.48 &N  &   (a)\\
J1424$+$4214 & 07.03.00 & ST  & R & 3.88 & 15 & 0.380 & 0.2 & 0.3  & 55.01  & 1.34 &N  &   (a)\\
J1424$+$4214 & 08.03.00 & GSO & V & 3.05 & 30 & 0.385 & 0.6 & 0.8  & 54.71  & 0.76 &N  &   (a)\\
&&&&&&&&&&&&\\
J1524$+$0958 & 11.04.99 & ST  & R & 6.55 & 38 & 0.491 & 0.2 & 0.3  & 78.81  & 0.96 &N  &   (a)\\
&&&&&&&&&&&&\\
J1528$+$2825 & 10.05.05 & ST  & R & 7.75 & 16 & 0.065 & 0.2 & 0.2  & 27.00  & 0.33 &N  &   (b)\\
&&&&&&&&&&&&\\
J1631$+$2953 & 15.06.04 & HCT & V & 6.21 & 28 & 1.110 & 0.2 & 0.4  & 64.31  & 1.00 &N  &   (b)\\
J1631$+$2953 & 11.05.05 & ST  & R & 6.92 & 29 & 0.006 & 0.3 & 0.4  & 53.36  & 0.62 &N  &   (b)\\
J1631$+$2953 & 01.06.05 & ST  & R & 7.36 & 15 & 1.369 & 0.2 & 0.4  & 30.35  & 0.93 &N  &   (b)\\
&&&&&&&&&&&&\\
J1632$+$3737 & 12.05.05 & ST  & R & 6.60 & 29 & 0.289 & 0.2 & 0.2  & 53.95  & 0.72 &N  &   (b)\\
&&&&&&&&&&&&\\
J1751$+$5045 & 03.06.98 & ST  & R & 4.72 & 46 & 0.373 & 0.2 & 0.3  & 109.29 & 1.00 &N  &   (a)\\
J1751$+$5045 & 06.06.98 & ST  & R & 1.65 & 17 & 0.384 & 0.3 & 0.4  & 32.15  & 0.93 &N  &   (a)\\
J1751$+$5045 & 08.06.98 & ST  & R & 6.15 & 36 & 0.021 & 0.2 & 0.3  & 157.64 & 1.78 &PV &   (a)\\
&&&&&&&&&&&&\\\hline
\multicolumn{13}{l}{\small Radio intermediate quasars (RIQs)[10 sources; 31 DLCs]}\\ \hline
J0005$+$1609 & 03.11.00 & ST  & R & 6.55 & 30 & 0.302 & 0.3 & 0.3  & 44.85  & 0.61 &N  &   (a)\\
J0005$+$1609 & 05.11.00 & ST  & R & 7.74 & 39 & 0.028 & 0.4 & 0.3  & 28.94  & 0.30 &N  &   (a)\\
&&&&&&&&&&&&\\
J0748$+$2200 & 19.01.07 & ST  & R & 5.20 & 19 & 0.030 & 0.3 & 0.3  & 28.12  & 0.62 &N  &   (d)\\
J0748$+$2200 & 23.01.07 & ST  & R & 7.21 & 25 & 0.149 & 0.3 & 0.4  & 38.97  & 0.64 &N  &   (d)\\
J0748$+$2200 & 19.02.07 & ST  & R & 6.42 & 24 & 0.614 & 0.3 & 0.4  & 77.17  & 1.24 &N  &   (d)\\
J0748$+$2200 & 29.01.08 & IGO & R & 5.41 & 19 & 0.627 & 0.1 & 0.1  & 17.96  & 0.42 &N  &   (d)\\
J0748$+$2200 & 30.01.08 & IGO & R & 6.03 & 20 & 0.805 & 0.1 & 0.2  & 33.25  & 0.67 &N  &   (d)\\
&&&&&&&&&&&&\\
J0832$+$3707 & 23.01.07 & HCT & R & 4.91 & 29 & 0.265 & 0.2 & 0.2  & 60.16  & 0.88 &N  &   (d)\\
J0832$+$3707 & 21.02.07 & ST  & R & 4.70 & 21 & 0.193 & 0.1 & 0.2  & 43.88  & 0.92 &N  &   (d)\\
J0832$+$3707 & 10.03.07 & IGO & R & 5.04 & 10 & 0.203 & 0.2 & 0.2  & 11.06  & 0.59 &N  &   (d)\\
J0832$+$3707 & 11.03.07 & IGO & R & 5.09 & 10 & 0.204 & 0.2 & 0.3  & 23.95  & 1.16 &N  &   (d)\\
&&&&&&&&&&&&\\
J0836$+$4426 & 22.01.07 & ST  & R & 5.61 & 24 & 1.288 & 0.2 & 0.2  & 19.63  & 0.35 &N  &   (d)\\
J0836$+$4426 & 10.02.07 & IGO & R & 5.58 & 15 & 0.815 & 0.2 & 0.3  & 36.26  & 1.00 &N  &   (d)\\
J0836$+$4426 & 09.03.07 & IGO & R & 5.16 & 16 & 0.864 & 0.2 & 0.3  & 39.16  & 1.49 &N  &   (d)\\
&&&&&&&&&&&&\\
J0907$+$5515 & 04.02.08 & IGO & R & 8.99 & 24 & 0.247 & 0.2 & 0.3  & 47.80  & 0.75 &N  &   (d)\\
J0907$+$5515 & 05.02.08 & IGO & R & 7.48 & 13 & 0.365 & 0.1 & 0.3  & 40.08  & 1.33 &N  &   (d)\\
&&&&&&&&&&&&\\
J1259$+$3423 & 19.04.07 & ST  & R & 5.40 & 21 & 0.673 & 0.2 & 0.4  & 95.09  & 1.63 &N  &   (d)\\
J1259$+$3423 & 20.04.07 & ST  & R & 6.40 & 27 & 0.673 & 0.2 & 0.3  & 66.00  & 0.80 &N  &   (d)\\
J1259$+$3423 & 24.04.07 & ST  & R & 5.30 & 22 & 0.688 & 0.2 & 0.3  & 41.81  & 0.79 &N  &   (d)\\
&&&&&&&&&&&&\\
J1312$+$3515 & 25.03.99 & ST  & R & 6.67 & 39 & 0.097 & 0.2 & 0.5  & 398.57 & 2.79 &V  &   (e)\\
J1312$+$3515 & 01.04.01 & ST  & R & 4.87 & 32 & 0.443 & 0.2 & 0.4  & 149.98 & 2.52 &V  &   (e)\\
J1312$+$3515 & 02.04.01 & ST  & R & 5.19 & 41 & 0.696 & 0.3 & 0.4  & 86.44  & 0.81 &N  &   (e)\\
&&&&&&&&&&&&\\
J1336$+$1725 & 11.04.05 & ST  & R & 7.93 & 29 & 0.305 & 0.1 & 0.2  & 53.60  & 0.80 &N  &   (d)\\
J1336$+$1725 & 08.05.05 & ST  & R & 4.47 & 17 & 0.739 & 0.2 & 0.3  & 60.18  & 1.53 &N  &   (d)\\
J1336$+$1725 & 13.04.08 & ST  & R & 8.06 & 20 & 0.731 & 0.2 & 0.3  & 56.65  & 1.33 &N  &   (d)\\
&&&&&&&&&&&&\\
J1539$+$4735 & 27.05.09 & ST  & R & 6.26 & 30 & 0.776 & 0.3 & 0.4  & 52.69  & 0.69 &N  &   (d)\\
J1539$+$4735 & 02.06.09 & ST  & R & 7.03 & 30 & 0.779 & 0.4 & 0.5  & 56.11  & 0.68 &N  &   (d)\\
J1539$+$4735 & 14.06.09 & ST  & R & 5.30 & 24 & 0.776 & 0.4 & 0.5  & 36.33  & 0.54 &N  &   (d)\\
&&&&&&&&&&&&\\
J1719$+$4804 & 29.04.06 & ST  & R & 4.88 & 25 & 0.131 & 0.1 & 0.2  & 54.32  & 0.95 &N  &   (d)\\
J1719$+$4804 & 30.04.06 & ST  & R & 5.64 & 22 & 0.195 & 0.1 & 0.2  & 61.02  & 1.22 &N  &   (d)\\
J1719$+$4804 & 30.05.06 & ST  & R & 6.06 & 26 & 0.031 & 0.2 & 0.3  & 62.64  & 0.85 &N  &   (d)\\
&&&&&&&&&&&&\\\hline
\multicolumn{13}{l}{\small Lobe dominated quasars (LDQs)[9 sources; 25 DLCs]}\\ \hline
J0015$+$3052 & 18.01.01 & ST  & R & 3.78 & 18 & 0.241 & 0.5 & 0.5  & 21.29  & 0.40 &N  &  (c) \\
J0015$+$3052 & 20.01.01 & ST  & R & 2.70 & 12 & 0.457 & 0.6 & 0.3  & 4.66   & 0.16 &N  &  (c) \\
J0015$+$3052 & 24.01.01 & ST  & R & 2.87 & 14 & 0.242 & 0.6 & 0.5  & 9.82   & 0.25 &N  &  (c) \\
J0015$+$3052 & 14.10.01 & ST  & R & 6.78 & 26 & 0.235 & 0.6 & 0.7  & 37.85  & 0.51 &N  &  (c) \\
J0015$+$3052 & 21.10.01 & ST  & R & 6.25 & 24 & 0.703 & 0.5 & 0.5  & 17.98  & 0.36 &N  &  (c) \\
&&&&&&&&&&&&\\
J0028$+$3103 & 13.10.98 & ST  & R & 3.60 & 28 & 0.241 & 0.1 & 0.2  & 57.87  & 0.90 &N  &  (a) \\
J0028$+$3103 & 01.11.98 & ST  & R & 3.35 & 26 & 0.260 & 0.2 & 0.3  & 76.98  & 1.14 &N  &  (a) \\
&&&&&&&&&&&&\\
J0137$+$3309 & 07.11.01 & ST  & R & 6.54 & 36 & 0.089 & 0.6 & 0.5  & 88.24  & 0.28 &N  &  (c) \\
J0137$+$3309 & 08.11.01 & ST  & R & 6.66 & 32 & 0.132 & 0.3 & 0.4  & 58.61  & 0.70 &N  &  (c) \\
J0137$+$3309 & 13.11.01 & ST  & R & 8.63 & 46 & 0.213 & 0.3 & 0.4  & 119.10 & 1.07 &N  &  (c) \\
&&&&&&&&&&&&\\
J0352$-$0711 & 14.11.01 & ST  & R & 6.56 & 31 & 0.617 & 0.2 & 0.3  & 70.99  & 0.80 &N  &  (c) \\
J0352$-$0711 & 15.11.01 & ST  & R & 5.54 & 26 & 0.630 & 0.2 & 0.3  & 39.30  & 0.66 &N  &  (c) \\
J0352$-$0711 & 18.11.01 & ST  & R & 5.70 & 25 & 0.628 & 0.2 & 0.4  & 106.55 & 1.42 &N  &  (c) \\
&&&&&&&&&&&&\\
J0713$+$3656 & 20.01.01 & ST  & R & 6.51 & 29 & 0.191 & 0.3 & 0.3  & 45.90  & 0.72 &N  &  (c) \\
J0713$+$3656 & 21.01.01 & ST  & R & 6.40 & 30 & 0.190 & 0.3 & 0.3  & 42.60  & 0.61 &N  &  (c) \\
J0713$+$3656 & 25.01.01 & ST  & R & 7.08 & 31 & 0.453 & 0.3 & 0.3  & 46.97  & 0.66 &N  &  (c) \\
J0713$+$3656 & 20.12.01 & ST  & R & 8.07 & 52 & 0.202 & 0.3 & 0.6  & 190.47 & 1.56 &N  &  (c) \\
J0713$+$3656 & 21.12.01 & ST  & R & 7.49 & 48 & 0.449 & 0.2 & 0.4  & 142.06 & 1.20 &N  &  (c) \\
&&&&&&&&&&&&\\
J1007$+$1248 & 16.02.99 & ST  & R & 6.51 & 36 & 1.000 & 0.1 & 0.3  & 213.36 & 2.42 &V  &  (c) \\
J1007$+$1248 & 27.02.99 & ST  & R & 4.27 & 30 & 0.996 & 0.4 & 0.4  & 39.70  & 0.51 &N  &  (c) \\
J1007$+$1248 & 29.03.00 & ST  & R & 3.81 & 21 & 1.012 & 0.1 & 0.2  & 58.34  & 1.23 &N  &  (c) \\
J1007$+$1248 & 30.03.00 & ST  & R & 4.64 & 26 & 1.007 & 0.2 & 0.3  & 71.58  & 0.83 &N  &  (c) \\
J1007$+$1248 & 18.02.01 & ST  & R & 5.54 & 42 & 1.015 & 0.2 & 0.4  & 112.96 & 1.16 &N  &  (c) \\
J1007$+$1248 & 24.03.01 & ST  & R & 6.38 & 50 & 1.011 & 0.2 & 0.4  & 297.51 & 1.91 &PV &  (c) \\
&&&&&&&&&&&&\\
J1106$-$0052 & 17.03.99 & ST  & R & 3.81 & 23 & 0.347 & 0.3 & 0.5  & 65.59  & 1.23 &N  &  (c) \\
J1106$-$0052 & 18.03.99 & ST  & R & 7.51 & 42 & 0.348 & 0.3 & 0.5  & 107.03 & 0.99 &N  &  (c) \\
J1106$-$0052 & 16.04.00 & ST  & R & 3.85 & 15 & 0.348 & 0.3 & 0.4  & 36.16  & 0.78 &N  &  (c) \\
J1106$-$0052 & 25.03.01 & ST  & R & 7.18 & 28 & 0.343 & 0.3 & 0.4  & 49.79  & 0.70 &N  &  (c) \\
J1106$-$0052 & 14.04.01 & ST  & R & 4.55 & 19 & 0.346 & 0.3 & 0.5  & 86.90  & 1.50 &N  &  (c) \\
J1106$-$0052 & 22.03.02 & ST  & R & 6.13 & 18 & 0.342 & 0.2 & 0.3  & 32.21  & 0.78 &N  &  (c) \\
&&&&&&&&&&&&\\
J1633$+$3924 & 04.06.99 & ST  & R & 5.71 & 30 & 0.293 & 0.6 & 0.6  & 28.75  & 0.45 &N  &  (a) \\
J1633$+$3924 & 30.05.00 & ST  & R & 3.54 & 14 & 0.542 & 0.5 & 0.6  & 15.95  & 0.52 &N  &  (a) \\
&&&&&&&&&&&&\\
J2351$-$0109 & 13.10.01 & ST  & R & 7.56 & 41 & 0.163 & 0.2 & 0.4  & 213.75 & 1.43 &N  &  (c) \\
J2351$-$0109 & 17.10.01 & ST  & R & 7.80 & 43 & 0.032 & 0.2 & 0.3  & 153.36 & 1.17 &N  &  (c) \\
J2351$-$0109 & 18.10.01 & ST  & R & 8.40 & 46 & 0.032 & 0.2 & 0.2  & 96.62  & 0.72 &N  &  (c) \\
&&&&&&&&&&&&\\\hline
\multicolumn{13}{l}{\small Low optical polarization core dominated quasars (LPCDQs)[12 sources; 43 DLCs]}\\ \hline
J0005$+$0524 & 23.10.06 & ST  & R & 7.05 & 16 & 0.132 & 0.3 & 0.2  & 11.64  & 0.31 &N  &  (f) \\
J0005$+$0524 & 18.11.06 & ST  & R & 4.69 & 11 & 0.394 & 0.2 & 0.1  & 6.30   & 0.24 &N  &  (f) \\
J0005$+$0524 & 14.09.07 & ST  & R & 5.31 & 12 & 0.370 & 0.2 & 0.4  & 30.33  & 1.14 &N  &  (f) \\
J0005$+$0524 & 16.09.07 & ST  & R & 6.11 & 13 & 0.240 & 0.2 & 0.4  & 81.99  & 2.15 &N  &  (f) \\
&&&&&&&&&&&&\\
J0235$-$0402 & 21.10.04 & ST  & R & 7.25 & 15 & 0.127 & 0.1 & 0.2  & 43.88  & 1.15 &N  &  (f) \\
J0235$-$0402 & 22.10.04 & ST  & R & 7.87 & 17 & 0.244 & 0.2 & 0.2  & 43.75  & 0.82 &N  &  (f) \\
J0235$-$0402 & 04.11.04 & ST  & R & 6.19 & 25 & 0.249 & 0.2 & 0.2  & 36.34  & 0.51 &N  &  (f) \\
J0235$-$0402 & 05.11.04 & ST  & R & 7.27 & 29 & 0.122 & 0.1 & 0.2  & 68.37  & 1.01 &N  &  (f) \\
&&&&&&&&&&&&\\
J0456$+$0400 & 23.11.08 & ST  & R & 5.50 & 24 & 0.405 & 0.2 & 0.3  & 43.41  & 0.79 &N  &  (f) \\
J0456$+$0400 & 29.11.08 & ST  & R & 5.51 & 20 & 0.404 & 0.2 & 0.3  & 36.82  & 0.82 &N  &  (f) \\
J0456$+$0400 & 03.12.08 & ST  & R & 5.38 & 22 & 0.529 & 0.3 & 0.3  & 28.65  & 0.59 &N  &  (f) \\
&&&&&&&&&&&&\\
J0741$+$3112 & 20.01.06 & ST  & R & 7.42 & 31 & 0.614 & 0.2 & 0.3  & 78.51  & 0.94 &N  &  (f) \\
J0741$+$3112 & 21.01.06 & ST  & R & 4.01 & 18 & 0.766 & 0.2 & 0.3  & 26.33  & 0.63 &N  &  (f) \\
J0741$+$3112 & 18.12.06 & ST  & R & 7.24 & 29 & 0.135 & 0.1 & 0.2  & 95.05  & 1.42 &N  &  (f) \\
J0741$+$3112 & 22.12.06 & ST  & R & 7.72 & 32 & 0.140 & 0.1 & 0.2  & 58.35  & 0.79 &N  &  (f) \\
&&&&&&&&&&&&\\
J0842$+$1835 & 04.02.06 & ST  & R & 7.64 & 28 & 0.274 & 0.1 & 0.2  & 59.41  & 0.92 &N  &  (f) \\
J0842$+$1835 & 16.12.06 & ST  & R & 5.96 & 14 & 0.277 & 0.1 & 0.4  & 83.30  & 2.57 &N  &  (f) \\
J0842$+$1835 & 21.12.06 & ST  & R & 6.94 & 30 & 0.279 & 0.1 & 0.2  & 92.31  & 1.23 &N  &  (f) \\
&&&&&&&&&&&&\\
J0958$+$3224 & 19.02.99 & ST  & R & 6.50 & 36 & 1.729 & 0.4 & 0.4  & 35.19  & 0.39 &N  &  (e) \\
J0958$+$3224 & 03.03.00 & ST  & R & 6.29 & 37 & 1.311 & 0.3 & 0.4  & 90.04  & 0.82 &N  &  (e) \\
J0958$+$3224 & 05.03.00 & ST  & R & 6.90 & 34 & 0.430 & 0.1 & 0.3  & 115.79 & 1.48 &N  &  (e) \\
&&&&&&&&&&&&\\
J1131$+$3114 & 18.01.01 & ST  & R & 5.73 & 31 & 0.230 & 0.3 & 0.4  & 59.11  & 0.83 &N  &  (e) \\
J1131$+$3114 & 09.03.02 & ST  & R & 8.22 & 27 & 0.435 & 0.3 & 0.3  & 41.50  & 0.52 &N  &  (e) \\
J1131$+$3114 & 10.03.02 & ST  & R & 8.33 & 28 & 0.200 & 0.2 & 0.3  & 46.60  & 0.66 &N  &  (e) \\
&&&&&&&&&&&&\\
J1228$+$3128 & 07.03.99 & ST  & R & 6.63 & 49 & 1.299 & 0.3 & 0.6  & 165.97 & 1.42 &N  &  (e) \\
J1228$+$3128 & 07.04.00 & ST  & R & 7.32 & 26 & 1.320 & 0.2 & 0.6  & 150.61 & 2.35P&V  &  (e) \\
J1228$+$3128 & 20.04.01 & ST  & R & 7.43 & 34 & 1.357 & 0.6 & 0.7  & 46.63  & 0.59 &N  &  (e) \\
&&&&&&&&&&&&\\
J1229$+$0203 & 07.03.11 & ST  & R & 5.46 & 35 & 0.084 & 0.1 & 0.2  & 61.36  & 0.72 &N  &  (f) \\
J1229$+$0203 & 10.03.11 & ST  & R & 6.72 & 49 & 0.047 & 0.1 & 0.2  & 114.60 & 1.00 &N  &  (f) \\
&&&&&&&&&&&&\\
J1357$+$1919 & 27.02.06 & ST  & R & 5.19 & 12 & 0.004 & 0.1 & 0.3  & 45.60  & 1.74 &N  &  (f) \\
J1357$+$1919 & 05.03.06 & ST  & R & 4.94 & 11 & 0.766 & 0.1 & 0.2  & 25.52  & 1.07 &N  &  (f) \\
J1357$+$1919 & 26.03.06 & ST  & R & 6.98 & 12 & 0.025 & 0.1 & 0.5  & 124.20 & 4.76 &V  &  (f) \\
J1357$+$1919 & 28.03.06 & ST  & R & 5.83 & 21 & 0.026 & 0.2 & 0.4  & 110.35 & 2.26P&V  &  (f) \\
J1357$+$1919 & 29.03.06 & ST  & R & 6.26 & 23 & 0.030 & 0.2 & 0.3  & 110.04 & 1.66 &N  &  (f) \\
J1357$+$1919 & 06.04.06 & ST  & R & 7.40 & 27 & 0.746 & 0.2 & 0.3  & 97.85  & 1.28 &N  &  (f) \\
J1357$+$1919 & 22.04.06 & ST  & R & 4.88 & 17 & 0.037 & 0.2 & 0.4  & 44.72  & 1.04 &N  &  (f) \\
J1357$+$1919 & 23.04.06 & ST  & R & 6.04 & 19 & 0.060 & 0.3 & 0.6  & 95.00  & 1.88 &N  &  (f) \\
&&&&&&&&&&&&\\
J2203$+$3145 & 08.11.05 & HCT & R & 5.62 & 18 & 0.478 & 0.2 & 0.3  & 92.02  & 1.38 &N  &  (f) \\
J2203$+$3145 & 14.09.06 & ST  & R & 5.87 & 26 & 0.158 & 0.2 & 0.3  & 78.55  & 1.27 &N  &  (f) \\
J2203$+$3145 & 15.09.07 & ST  & R & 7.74 & 33 & 0.511 & 0.2 & 0.2  & 38.25  & 0.75 &N  &  (f) \\
&&&&&&&&&&&&\\
J2346$+$0930 & 20.09.03 & HCT & R & 5.82 & 39 & 0.772 & 0.1 & 0.3  & 137.92 & 1.65 &N  &  (f) \\
J2346$+$0930 & 20.10.04 & ST  & R & 5.73 & 11 & 0.128 & 0.1 & 0.3  & 52.59  & 2.21 &N  &  (f) \\
J2346$+$0930 & 16.11.06 & ST  & R & 5.24 & 12 & 0.732 & 0.2 & 0.2  & 18.58  & 0.68 &N  &  (f) \\
&&&&&&&&&&&&\\\hline
\multicolumn{13}{l}{\small High optical polarization core dominated quasars (HPCDQs)[11 sources; 31 DLCs]}\\ \hline
J0238$+$1637 & 12.11.99 & ST  & R & 6.57 & 40 & 1.016 & 0.4 & 0.7  & 95.28  & 1.08 &N  &  (e) \\
J0238$+$1637 & 14.11.99 & ST  & R & 6.16 & 34 & 1.020 & 0.2 & 0.4  & 88.31  & 1.13 &N  &  (e) \\
J0238$+$1637 & 18.11.03 & HCT & R & 7.80 & 41 & 0.251 & 0.3 & 0.5  & 129.42 & 1.34 &N  &  (f) \\
&&&&&&&&&&&&\\
J0423$-$0120 & 19.11.03 & HCT & R & 6.69 & 38 & 0.402 & 0.2 & 0.3  & 153.68 & 1.41 &N  &  (f) \\
J0423$-$0120 & 08.12.04 & ST  & R & 7.00 & 13 & 0.412 & 0.1 & 0.3  & 38.52  & 1.21 &N  &  (f) \\
J0423$-$0120 & 25.10.09 & ST  & R & 4.46 & 21 & 0.128 & 0.3 & 0.6  & 76.55  & 1.48 &N  &  (f) \\
&&&&&&&&&&&&\\
J0739$+$0137 & 05.12.05 & HCT & R & 5.31 & 10 & 0.461 & 0.1 & 0.2  & 20.17  & 0.94 &N  &  (f) \\
J0739$+$0137 & 06.12.05 & HCT & R & 6.06 & 9  & 0.647 & 0.1 & 0.4  & 80.48  & 4.24 &PV &  (f) \\
J0739$+$0137 & 09.12.05 & HCT & R & 5.46 & 14 & 0.186 & 0.1 & 0.3  & 57.77  & 1.87 &N  &  (f) \\
&&&&&&&&&&&&\\
J0849$+$5108 & 30.12.98 & ST  & R & 7.08 & 39 & 0.603 & 0.8 & 1.3  & 116.19 & 1.18 &N  &  (a) \\
&&&&&&&&&&&&\\
J1058$+$0133 & 25.03.07 & ST  & R & 6.87 & 13 & 0.177 & 0.1 & 0.2  & 21.51  & 0.81 &N  &  (f) \\
J1058$+$0133 & 16.04.07 & ST  & R & 4.23 & 17 & 0.501 & 0.1 & 0.2  & 52.55  & 1.38 &N  &  (f) \\
J1058$+$0133 & 23.04.07 & ST  & R & 5.36 & 12 & 0.158 & 0.2 & 0.3  & 25.12  & 0.81 &N  &  (f) \\
&&&&&&&&&&&&\\
J1159$+$2914 & 31.03.12 & IGO & R & 5.93 & 18 & 0.134 & 0.6 & 0.7  & 34.89  & 0.53 &N  &  (f) \\
J1159$+$2914 & 01.04.12 & IGO & R & 8.40 & 26 & 0.133 & 0.8 & 0.9  & 39.13  & 0.61 &N  &  (f) \\
J1159$+$2914 & 02.04.12 & IGO & R & 7.22 & 20 & 0.144 & 1.5 & 2.9  & 69.58  & 1.59 &N  &  (f) \\
&&&&&&&&&&&&\\
J1218$-$0119 & 11.03.02 & ST  & R & 6.16 & 34 & 0.049 & 1.3 & 3.0  & 225.39 & 2.39 &PV &  (e) \\
J1218$-$0119 & 13.03.02 & ST  & R & 8.48 & 24 & 0.074 & 0.2 & 0.5  & 158.12 & 1.62 &N  &  (e) \\
J1218$-$0119 & 15.03.02 & ST  & R & 3.91 & 11 & 0.077 & 0.2 & 0.3  & 29.68  & 0.59 &N  &  (e) \\
J1218$-$0119 & 16.03.02 & ST  & R & 8.20 & 22 & 0.072 & 0.2 & 0.3  & 121.40 & 1.52 &N  &  (e) \\
&&&&&&&&&&&&\\
J1256$-$0547 & 26.01.06 & ST  & R & 4.75 & 21 & 0.596 & 0.1 & 0.2  & 65.87  & 1.38 &N  &  (f) \\
J1256$-$0547 & 28.02.06 & ST  & R & 6.51 & 42 & 0.601 & 0.1 & 0.2  & 91.54  & 0.81 &N  &  (f) \\
J1256$-$0547 & 20.04.09 & ST  & R & 5.46 & 22 & 0.601 & 0.2 & 0.3  & 43.51  & 0.75 &N  &  (f) \\
&&&&&&&&&&&&\\
J1310$+$3220 & 26.04.00 & ST  & R & 5.99 & 18 & 0.971 & 1.0 & 1.8  & 48.79  & 1.34 &N  &  (e) \\
J1310$+$3220 & 17.03.02 & ST  & R & 8.37 & 21 & 1.050 & 0.8 & 0.6  & 17.49  & 0.27 &N  &  (e) \\
J1310$+$3220 & 24.04.02 & ST  & R & 5.81 & 14 & 1.045 & 0.5 & 0.3  & 7.70   & 0.17 &N  &  (e) \\
J1310$+$3220 & 02.05.02 & ST  & R & 5.08 & 15 & 0.031 & 0.5 & 0.4  & 8.61   & 0.21 &N  &  (e) \\
&&&&&&&&&&&&\\
J1512$-$0906 & 14.06.05 & ST  & R & 4.93 & 11 & 0.347 & 0.1 & 0.1  & 9.33   & 0.39 &N  &  (f) \\
J1512$-$0906 & 01.05.09 & ST  & R & 6.02 & 25 & 0.557 & 0.3 & 0.5  & 58.70  & 1.02 &N  &  (f) \\
J1512$-$0906 & 20.05.09 & ST  & R & 5.16 & 25 & 0.580 & 0.5 & 0.7  & 55.86  & 0.67 &N  &  (f) \\
&&&&&&&&&&&&\\
J2222$-$0457 & 08.10.10 & ST  & R & 5.72 & 18 & 0.044 & 0.4 & 0.9  & 69.00  & 1.59 &N  &  (g) \\
&&&&&&&&&&&&\\\hline
\multicolumn{13}{l}{\small TeV detected BL Lac objects (TeV-BLs)[13 sources; 54 DLCs]}\\ \hline
J0112$+$2244 & 29.10.05 & ST  & R & 7.14 & 36 & 0.250 & 0.1 & 0.2  & 71.03  & 0.85 &N  &  (h) \\
&&&&&&&&&&&&\\
J0222$+$4302 & 13.11.99 & ST  & R & 5.92 & 123 & 0.051 & 0.1& 0.2  & 416.3  & 1.43 &PV &  (i) \\
J0222$+$4302 & 24.10.00 & ST  & R & 9.15 & 73  & 0.050 & 0.1& 0.3  & 310.17 & 1.95 &V  &  (i) \\
J0222$+$4302 & 01.11.00 & ST  & R & 9.02 & 103 & 0.363 & 0.2& 0.3  & 218.47 & 0.86 &N  &  (i) \\
&&&&&&&&&&&&\\
J0721$+$7120 & 01.02.05 & ST  & R & 1.68 & 26 & 0.159 & 0.2 & 0.3  & 62.62  & 0.86 &N  &  (g) \\
&&&&&&&&&&&&\\
J0738$+$1742 & 26.12.98 & ST  & R & 7.79 & 49 & 0.122 & 0.4 & 0.6  & 89.48  & 0.75 &N  &  (j) \\
J0738$+$1742 & 30.12.99 & ST  & R & 7.44 & 64 & 0.066 & 0.4 & 0.5  & 96.90  & 0.64 &N  &  (j) \\
J0738$+$1742 & 25.12.00 & ST  & R & 6.01 & 42 & 0.061 & 0.4 & 0.5  & 69.02  & 0.69 &N  &  (j) \\
J0738$+$1742 & 24.12.01 & ST  & R & 7.30 & 38 & 0.190 & 0.3 & 0.4  & 47.70  & 0.52 &N  &  (j) \\
J0738$+$1742 & 20.12.03 & HCT & R & 6.00 & 38 & 0.818 & 0.2 & 0.3  & 71.02  & 0.80 &N  &  (j) \\
J0738$+$1742 & 10.12.04 & ST  & R & 6.23 & 30 & 0.512 & 0.2 & 0.3  & 98.67  & 1.17 &N  &  (j) \\
J0738$+$1742 & 23.12.04 & ST  & R & 5.88 & 13 & 0.505 & 0.1 & 0.2  & 36.57  & 1.15 &N  &  (j) \\
J0738$+$1742 & 02.01.05 & ST  & R & 4.87 & 22 & 0.522 & 0.2 & 0.2  & 29.93  & 0.81 &N  &  (j) \\
J0738$+$1742 & 05.01.05 & ST  & R & 5.23 & 26 & 0.158 & 0.1 & 0.2  & 64.56  & 1.08 &N  &  (j) \\
J0738$+$1742 & 09.01.05 & ST  & R & 7.13 & 30 & 0.152 & 0.1 & 0.2  & 64.47  & 0.90 &N  &  (j) \\
J0738$+$1742 & 09.11.05 & ST  & R & 4.27 & 19 & 0.624 & 0.1 & 0.2  & 48.34  & 1.13 &N  &  (j) \\
J0738$+$1742 & 16.11.06 & ST  & R & 4.97 & 21 & 0.033 & 0.2 & 0.3  & 64.94  & 1.10 &N  &  (j) \\
J0738$+$1742 & 29.11.06 & ST  & R & 6.49 & 28 & 0.516 & 0.2 & 0.3  & 66.83  & 1.00 &N  &  (j) \\
J0738$+$1742 & 17.12.06 & ST  & R & 6.54 & 28 & 0.507 & 0.1 & 0.3  & 118.30 & 1.45 &N  &  (j) \\
J0738$+$1742 & 15.12.07 & ST  & R & 7.05 & 29 & 0.162 & 0.1 & 0.2  & 89.88  & 1.35 &N  &  (j) \\
J0738$+$1742 & 16.12.07 & ST  & R & 7.29 & 30 & 0.508 & 0.2 & 0.2  & 30.66  & 0.42 &N  &  (j) \\
J0738$+$1742 & 22.11.08 & ST  & R & 5.98 & 29 & 0.128 & 0.2 & 0.2  & 48.35  & 0.53 &N  &  (j) \\
J0738$+$1742 & 08.12.09 & ST  & R & 6.94 & 31 & 0.128 & 0.3 & 0.5  & 80.87  & 0.91 &N  &  (h) \\
J0738$+$1742 & 05.01.11 & ST  & R & 6.80 & 32 & 0.330 & 0.3 & 0.4  & 43.17  & 0.51 &N  &  (h) \\
J0738$+$1742 & 29.11.11 & ST  & R & 6.11 & 29 & 0.499 & 0.2 & 0.3  & 34.25  & 0.51 &N  &  (h) \\
&&&&&&&&&&&&\\
J0809$+$3122 & 28.12.98 & ST  & R & 7.29 & 36 & 0.844 & 0.3 & 0.6  & 153.04 & 1.69 &N  &  (a) \\
&&&&&&&&&&&&\\
J0809$+$5218 & 04.02.05 & HCT & R & 7.24 & 29 & 0.885 & 0.1 & 0.3  & 97.92  & 1.43 &N  &  (g) \\
J0809$+$5218 & 05.12.05 & HCT & R & 5.85 & 10 & 0.892 & 0.1 & 0.3  & 31.21  & 1.26 &N  &  (g) \\
J0809$+$5218 & 08.12.05 & HCT & R & 5.77 & 16 & 0.894 & 0.2 & 0.2  & 18.25  & 0.40 &N  &  (g) \\
J0809$+$5218 & 09.12.05 & HCT & R & 5.46 & 14 & 0.892 & 0.2 & 0.2  & 17.38  & 0.56 &N  &  (g) \\
&&&&&&&&&&&&\\
J0854$+$2006 & 29.12.98 & ST  & R & 6.77 & 19 & 0.014 & 1.0 & 0.5  & 4.27   & 0.10 &N  &   (i)\\
J0854$+$2006 & 31.12.99 & ST  & R & 5.61 & 29 & 0.471 & 0.2 & 0.4  & 98.30  & 1.48 &N  &   (i)\\
J0854$+$2006 & 28.03.00 & ST  & R & 4.24 & 22 & 0.462 & 0.4 & 0.5  & 29.78  & 0.64 &N  &   (i)\\
J0854$+$2006 & 17.02.01 & ST  & R & 6.92 & 47 & 0.467 & 0.4 & 0.4  & 46.55  & 0.42 &N  &   (i)\\
J0854$+$2006 & 05.02.05 & HCT & R & 7.82 & 42 & 1.739 & 0.1 & 0.2  & 127.8  & 1.05 &N  &   (g)\\
J0854$+$2006 & 12.04.05 & ST  & R & 4.77 & 56 & 0.907 & 0.3 & 0.4  & 65.20  & 0.45 &N  &   (g)\\
&&&&&&&&&&&&\\
J1015$+$4926 & 06.02.10 & ST  & R & 5.93 & 26 & 0.248 & 0.1 & 0.2  & 84.52  & 1.42 &N  &  (g) \\
J1015$+$4926 & 19.02.10 & ST  & R & 6.05 & 43 & 0.252 & 0.2 & 0.3  & 171.66 & 1.26 &N  &  (g) \\
J1015$+$4926 & 07.03.10 & ST  & R & 5.50 & 36 & 0.180 & 0.2 & 0.4  & 132.23 & 1.14 &N  &  (g) \\
&&&&&&&&&&&&\\
J1221$+$2813 & 19.03.04 & ST  & R & 6.20 & 60 & 2.324 & 0.3 & 0.5  & 159.14 & 1.14 &N  &  (g) \\
J1221$+$2813 & 20.03.04 & ST  & R & 6.29 & 67 & 2.322 & 0.4 & 0.7  & 196.68 & 1.08 &N  &  (g) \\
J1221$+$2813 & 18.03.05 & ST  & R & 4.18 & 28 & 1.301 & 0.2 & 0.5  & 116.81 & 2.22 &PV &  (g) \\
J1221$+$2813 & 05.04.05 & ST  & R & 7.28 & 41 & 1.280 & 0.2 & 0.4  & 170.26 & 1.75 &PV &  (g) \\
&&&&&&&&&&&&\\
J1221$+$3010 & 08.03.10 & IGO & R & 6.54 & 17 & 0.004 & 0.1 & 0.4  & 123.33 & 2.84 &PV &  (g) \\
J1221$+$3010 & 18.03.10 & ST  & R & 5.87 & 27 & 1.016 & 0.3 & 0.4  & 41.95  & 0.70 &N  &  (g) \\
J1221$+$3010 & 22.05.10 & ST  & R & 4.21 & 21 & 0.009 & 1.3 & 1.4  & 25.99  & 0.50 &N  &  (g) \\
&&&&&&&&&&&&\\
J1419$+$5423 & 28.03.99 & ST  & R & 5.65 & 33 & 0.142 & 0.3 & 0.5  & 68.98  & 0.82 &N  &  (a) \\
&&&&&&&&&&&&\\
J1428$+$4240 & 21.04.04 & HCT & R & 6.12 & 35 & 0.865 & 0.4 & 0.8  & 165.94 & 1.54 &N  &  (g) \\
J1428$+$4240 & 22.04.09 & ST  & R & 4.48 & 19 & 0.306 & 0.6 & 0.8  & 28.34  & 0.72 &N  &  (g) \\
J1428$+$4240 & 29.04.09 & ST  & R & 6.81 & 29 & 0.856 & 0.6 & 0.9  & 78.27  & 0.86 &N  &  (g) \\
&&&&&&&&&&&&\\
J1555$+$1111 & 05.05.99 & ST  & R & 4.15 & 23 & 1.170 & 0.3 & 0.5  & 65.67  & 1.26 &N  &   (a)\\
J1555$+$1111 & 24.06.09 & ST  & R & 4.22 & 26 & 0.137 & 0.1 & 0.3  & 108.25 & 1.77 &N  &   (g)\\
J1555$+$1111 & 15.05.10 & ST  & R & 6.50 & 22 & 0.041 & 0.1 & 0.3  & 112.32 & 1.98 &N  &   (g)\\
J1555$+$1111 & 16.05.10 & ST  & R & 6.27 & 33 & 0.101 & 0.2 & 0.3  & 164.16 & 1.53 &N  &   (g)\\
&&&&&&&&&&&&\\
\enddata

{\small
Columns :- (1) source name; (2) date of observation; (3) telescope used; (4) filter used; (5) duration of
monitoring; (6) number of data points in the DLC; (7) mean apparent magnitude difference of the steady star-star pair;
(8) quadratic mean of the $\sc IRAF$ errors for the steady star-star DLC;
(9) standard deviation of the steady star-star DLC;
(10) $\chi^2$-value for the star-star DLC; (10) $F$-value for the star-star DLC;
(12) variability status for the star-star DLC;
(13) reference for the INOV data.\\

{$^\P$} ST - Sampurnanand Telescope (ARIES); HCT - Himalayan Chandra Telescope (IIA);
IGO - IUCAA Girawali Observatory; VBT - Vainu Bappu Telescope (IIA);  GSO - Gurushikhar telecsope (PRL). \\
$^\dag$ V = Variable; N = Non-variable; PV = Probable Variable; \\
{$^\pounds$}References for the INOV data:
(a) \citet{2005MNRAS.356..607Stalin}; (b) \citet{2007BASI...35..141AG};
(c) \citet{2004MNRAS.350..175Stalin}; (d) \citet{2010MNRAS.401.2622AG};
(e) \citet{2004MNRAS.348..176Sagar}; (f) \citet{2012A&A...544A..37AG};
(g) \citet{2011MNRAS.416..101GK}; (h) AGs unpublished data;
(i) \citet{2004JApA...25....1Stalin}; (j) \citet{2009MNRAS.399.1622AG}.
}

\end{deluxetable}
\section*{Acknowledgements}
AG would like to thank Dr. Santosh Joshi (ARIES) for carrring out 
optical observations on a few occasions. The authors are thankful 
to the anonymous referee for the critical and constructive
suggestions.
\\
\bibliographystyle{mn2e}
\bibliography{mybib}
\end{document}